\documentstyle [11pt]{article}
\title{ {\bf Skewness as a Test of Non--Gaussian Primordial
Density Fluctuations}}
\author{{\bf Peter Coles}$^1$, {\bf Lauro Moscardini}$^{2,3}$,
{\bf Francesco Lucchin}$^2$,\\
{\bf Sabino Matarrese}$^4$ and  {\bf Antonio Messina}$^{5,6}$\\ \\
$^1$ {\it Astronomy Unit, School of Mathematical Sciences,} \\
{\it Queen Mary \& Westfield College, Mile End Road,} \\
{\it London E1 4NS, UK}\\ \\
$^2$ {\it Dipartimento  di Astronomia,}
{\it Universit\`{a} di Padova,}\\ {\it vicolo dell'Osservatorio 5,
I--35122 Padova, Italy}\\ \\
$^3$ {\it Astronomy Centre, University of Sussex,}\\
{\it Falmer, Brighton BN1 9QH, UK}\\ \\
$^4$ {\it Dipartimento di Fisica G. Galilei, Universit\`{a} di Padova,}\\
{\it via Marzolo 8, I--35131 Padova, Italy}\\ \\
$^5$ {\it Centre Europ\'een de Calcul Atomique et Moleculaire,}\\
{\it 91405 Orsay CEDEX, France}\\ \\
$^6$ {\it Dipartimento di Fisica A. Righi,  Universit\`{a} di Bologna,}\\
{\it via Irnerio 46, I--40126 Bologna, Italy}\\ \\}

\date{}

\begin{document}

\maketitle

\thispagestyle{empty}

\noindent{\bf Summary.} We investigate the evolution of the skewness of
the distribution of density fluctuations in CDM models with both
Gaussian and non--Gaussian initial fluctuations. We show that the
method proposed by Coles \& Frenk (1991), which uses the skewness
of galaxy counts to test the hypothesis of Gaussian primordial density
fluctuations, is a potentially powerful probe of initial conditions.
As expected, the mass distribution in models with initially non--Gaussian
fluctuations shows systematic
departures from the Gaussian behaviour on intermediate to large
scales. We investigate the effect of peculiar velocity distortions
and normalisation upon the relationship between skewness and variance.
These effects are generally small for the models we consider.
Comparing our results to the QDOT measurements
of the skewness, we find that our initially positive--skew models
are clearly excluded by this analysis, but the available data
do not rule out the negative--skew models.

\noindent{\bf Key Words:} Galaxies: formation, clustering -- large-scale
structure of the Universe -- early Universe -- dark matter.

\newpage

\setcounter{page}{1}
\section{Introduction}

The gravitational instability picture of galaxy formation is coming
under increasing pressure from new observational data. In particular,
models of structure formation involving dark matter -- either
hot (HDM) or cold (CDM) -- have been found wanting by new information on
galaxy clustering and the Cosmic Microwave Background Radiation (CMBR).
The ``standard'' versions of both HDM and CDM incorporate the
assumption that present--day structures grew by gravitational instability
from small, primordial, Gaussian distributed
adiabatic perturbations with the scale--invariant Zel'dovich spectrum.
Both these ``standard" cosmogonies seem however to be unable to account
for all the observational constraints. As far as galaxy clustering
data are concerned:
HDM succeeds in reproducing the amount of clustering on very large scales
but fails in accounting for the age of galaxies and galaxies must be
much less clustered than the mass on small scales to reproduce known
clustering properties; CDM on the other hand
better reproduces small--scale structures but suffers from general
lack of power on large scales. In particular, the large--scale problem of CDM
is indicated by a number of independent statistical tests such
as counts of galaxies in cells (Efstathiou {\em et al.} 1990;
Saunders {\em et al.} 1991), the spatial two--point correlation
function of rich
clusters (Batuski, Melott \& Burns 1987), the angular correlations of projected
galaxy distributions (Maddox {\em et al.} 1990) and the power--spectrum
of the distribution of galaxies and radio--galaxies (Peacock 1991).

In recent months, the COBE detection of large--scale temperature
anisotropy in the CMBR has given independent information about the
normalisation of fluctuations in these models, and the shape of the
power--spectrum on very large scales (Smoot {\em et al.} 1992).
This poses a particular problem for
CDM, since the normalisation implied by COBE is rather high for the
standard model. Normalised to match the COBE amplitude, CDM does
quite well on large scales but has too high an amplitude on small
scales, resulting in runaway clustering and very high peculiar
velocities (Davis {\em et al.} 1985). Attempts to escape
from these difficulties by a straightforward
change in either the amplitude or the shape of the primordial power
spectrum (e.g. Bardeen, Bond \& Efstathiou 1987; Vittorio, Matarrese \&
Lucchin 1988) are strongly constrained, not only by COBE but also
by other observational limits on temperature anisotropies on
the microwave background sky
(Bond {\em et al.} 1991; Vittorio {\em et al.} 1991). Nevertheless, it
is possible to beat the COBE constraints by invoking a power--law
inflationary model (Abbott \& Wise 1984, Lucchin \& Matarrese 1985a,b) that
produces almost, but not quite, scale--invariant
density fluctuations but also produces tensor perturbations (i.e.
gravitational waves) with large amplitude (Davis {\em et al.} 1992;
Liddle \& Lyth 1992; Lidsey \& Coles 1992; Lucchin, Matarrese \& Mollerach
1992; Salopek 1992b; Souradeep \& Sahni 1992). Such a model can allow a lower
amplitude of density fluctuations to be compatible with the COBE
anisotropy and can thus reconcile small--scale clustering with the CMBR
anisotropy.

Another possible escape route from these constraints is to assume that the
relationship between galaxies and mass is different on different mass
scales: a {\em scale--dependent bias}. Given how little we know about
galaxy formation this seems a reasonable choice and specific models have
been constructed that can alleviate the large--scale structure problem
for CDM without damaging its small--scale successes (Babul \& White 1991;
Bower {\em et al.} 1992). The problem with these models is that they
need to invoke collective physical processes acting on a very large
scale $\sim 30 h^{-1}$ Mpc and feedback mechanisms acting on such large
scales are hard to find. The need to introduce a physical scale into the
problem is demonstrated by Coles (1992): any {\em local} biasing effect
acting upon Gaussian density fluctuations
cannot change the slope of the galaxy correlation function with respect
to the mass autocorrelation function.

It is clear that many of the problems with large--scale structure theory
can be traced back to the assumption of random--phase fluctuations
within the gravitational instability model. Can we construct a
self--consistent model for galaxy formation based upon non--Gaussian
fluctuations?
Within the inflationary picture of the origin of
perturbations this issue has been discussed by a number of authors
(e.g. Matarrese, Ortolan \& Lucchin 1989; Kofman {\em et al.} 1991;
Barrow \& Coles 1990). The actual possibility of getting phase correlations on
cosmologically relevant scales is restricted to multiple scalar field models
(e.g. Allen, Grinstein \& Wise 1987; Salopek, Bond \& Bardeen 1989; Salopek
\& Bond 1991; Salopek 1992a; Fan \& Bardeen 1992). Alternatively,
non--Gaussian fluctuations can be produced by a
discrete, random distribution of seed masses, such as topological defects
like monopoles, cosmic strings or textures, provided by a phase--transition
in the early universe (Turok 1989; Park, Spergel \& Turok 1991;
Scherrer \& Bertschinger 1991; Scherrer 1992).
An alternative way to get non--random phases is the cosmic
explosion  scenario (Ostriker \& Cowie 1981; Ikeuchi 1981), where
hydrodynamics rather than gravity plays the main role in the structure
formation process.

A typical (though not mandatory) signature of non--Gaussian density
fluctuations\footnote{As usual, $\delta$ is defined to be the
dimensionless density contrast: $\delta=(\rho-\bar{\rho})/\bar{\rho}$,
where $\bar{\rho}$ is the mean matter density}, $\delta_M$,
is an initially non--vanishing
skewness $\langle \delta_M^3 \rangle \neq 0$. Actually, according to
the analysis of the QDOT IRAS--selected data by Saunders {\em et al.} (1991),
a positive skewness of the distribution of IRAS galaxy counts
is observed on quite large scales. A number of recent papers
debate the issue of whether such a positive
skewness has a primordial origin or is due simply to the non--linear,
aggregating action of gravity on a primordial (i.e. unskewed)
Gaussian field
(e.g. Coles \& Frenk 1991, hereafter CF91;
Silk \& Juszkiewicz 1991; Martel \& Freudling 1991; Park 1991;
Lahav {\em et al.} 1992; Juszkiewicz, Bouchet
\& Colombi 1992; Bouchet {\em et al.} 1992; Bouchet \& Hernquist 1992). It
has been shown by $N$--body simulations (Moscardini {\em et al.} 1991,
hereafter MMLM; Matarrese {\em et al.} 1991, Weinberg \& Cole 1992) that the
primordial skewness is a strongly discriminating
parameter in determining both the dynamics and the present clustering
properties of the universe. MMLM have studied
the origin of large--scale structures in skewed CDM models, while
Weinberg \& Cole (1992) have considered non--Gaussian
models obtained by a local non--linear transformation on scale--free
Gaussian fields. These studies also show that there are many similarities
between skew--positive
models and e.g. the texture--seeded CDM model (Park, Spergel \& Turok 1991;
Cen {\em et al.} 1991).
Actually, scenarios based on accreting HDM or CDM around seed
masses always induce an excess of overdense regions.
Cosmic explosions (e.g. Weinberg, Ostriker \& Dekel 1989), as well as
bubbles left over by a period
of extended inflation in the early universe (La \& Steinhardt 1989a,b;
Liddle \& Wands 1991), both
give rise to an excess of low density regions, resembling the initial
conditions of primordial skew--negative models.

So the {\em initial} skewness of non--Gaussian models is a strong
indicator of their clustering behaviour. But, as we have mentioned, this
primordial skewness is masked to some extent by the effect of
gravitational evolution which generally tends to couple the skewness to
the variance, which increases in time. Can we disentangle these two
possible causes of skewness and use the skewness of the present--day
distribution as a test of the hypothesis of Gaussian fluctuations?
Based on analytical arguments and $N$--body experiments,
CF91 argued that the answer to this question is
{\em yes} and they proposed a simple test for primordial
non--Gaussianity which is almost independent of the primordial spectral
index (Bouchet {\em et al.} 1992): it is the purpose of this paper
to determine how powerful
is the test suggested by CF91 against various non--Gaussian
alternatives. To this we have used the results of $N$--body
simulations with both Gaussian and non--Gaussian initial conditions
(Messina {\em et al.} 1992; Lucchin {\em et al.} in preparation)
which represent the universe on a cube of $260~h^{-1}$ Mpc side.
We have already used these simulations in a study of the large--scale
topology of the Universe (Coles {\em et al.} 1992, hereafter CMPLMM),
which is an alternative way of testing the Gaussian hypothesis
(for a review, see Melott 1990).

\section{$N$--body simulations with Non--Gaussian Initial Conditions}

The non--Gaussian statistics considered here are the same adopted by MMLM,
namely the {\em Lognormal} (hereafter LN) and the
{\em Chi--squared} with one degree of freedom (hereafter
$\chi^2$), chosen as distributions for the peculiar
gravitational potential, $\Phi$, before the modulation by the CDM
transfer function. These distributions actually split in two different
types of models: the {\em positive} ($LN_p$ and $\chi^2_p$)
and {\em negative} ($LN_n$ and $\chi^2_n$) models, classified
according to the sign of the skewness for linear mass fluctuations,
$\langle \delta_M^3\rangle>0$, for the former and
$\langle \delta_M^3\rangle<0$ for the latter.

We have to restrict the parameter space we analyse in some way, so
we set up all our model distributions in such a way that $\Phi$ has the
``standard" CDM power--spectrum
$$
{\cal P}_\Phi(k) = {9 \over 4} {\cal P}_0 k^{-3} T^2(k),
\eqno(2.1)
$$
where ${\cal P}_0$ is a normalization constant and $T(k)$ is the CDM transfer
function (e.g. Davis {\em et al.} 1985)
$$
T(k) = [1 + 6.8 k + 72.0 k^{3/2} + 16.0 k^2 ]^{-1},
\eqno(2.2)
$$
having considered a flat universe with Hubble constant $h=0.5$ in units
of $100$ Km sec$^{-1}$ Mpc$^{-1}$.
Such a choice for the spectrum allows a direct comparison with
the Gaussian CDM (hereafter G) model.

We used a particle--mesh code with $N_p = 128^3$ particles on
$N_g=128^3$ grid--points [more details will be given in a
forthcoming paper (Lucchin {\em et al.} in preparation)].
Computations were performed at the CINECA Center (Bologna) on a Cray YMP/432
running under UNICOS.
The box--size of our simulations is $L=260 ~h^{-1}$ Mpc;
each particle has mass $m = 4.7 \times 10^{12} {\rm M}_\odot$.
We run two realizations for each of the five models we consider.
We evolve our models starting from the same amplitude
up to the ``present time" $t_0$.
We define $t_0$ as the time when the galaxy two--point
function is best fitted by the power--law $\xi(r)=(r/r_0)^{-\gamma}$,
with $\gamma=1.8$ in a suitable interval.
To obtain the {\em galaxies} in a given simulation we proceed as follows.
We filter the initial density field with a Gaussian window function of radius
$0.5~h^{-1}$ Mpc and pick up as galaxies the particles
closest to each peak, defined as the grid--point with a positive density
contrast larger than the 26 nearest grid--points:
the result is a galaxy catalogue formed by $\sim 60,000$ per each
simulation. Due to the exceedingly high mass of our particles, following from
the large box size and low resolution, and to the rather simplified
galaxy identification criterion, we can only assume that
our peak regions roughly trace the actual galaxy distribution.

Different epochs can be parameterized by the {\em bias factor} $b$
defined by the {\em rms} linear mass--fluctuation
on a sharp--edged sphere of radius $R_8=8~h^{-1}$ Mpc, i.e.
$$
\sigma^2(R_8) = {{\cal P}_0 \over 2 \pi^2} \int_0^\infty dk k^3
T^2(k) W^2_{TH}(kR_8)  = {1\over b^2},
\eqno(2.3)
$$
where $W_{TH}(x)=(3/x)j_1(x)$ is a top--hat window function and
$j_1$ is the Bessel function of order $1$.
The present time $t_0$ corresponds to $b=1$ for the Gaussian model,
$b=1.5$ for both the positive models, $b=0.5$ for the negative $\chi^2$ and
$b=0.4$ for the negative Lognormal.
Note that the method to define the galaxies used in this work is different
from the 'excursion regions' technique used in CMPLMM, where a larger galaxy
number density, $3 \times 10^{-2}~h^3$ Mpc$^{-3}$, was necessary in order to
generate simulated Lick catalogs with $\sim 530,000$ galaxies in the whole
simulation box. A consequence of this change is, for example, that the present
epoch, i.e. the slope $\gamma=1.8$ for the correlation function, is reached
slightly later here than in CMPLMM.
Note that a fully consistent normalization of mass fluctuations should give
CMBR fluctuations in agreement with those detected by COBE (Smoot {\em et al.}
1992), which, for a standard CDM model favour low values of $b$,
namely $b \approx 0.8$. On the other hand, the statistical analysis of CMBR
anisotropies on large angular scales for non--Gaussian models cannot be
reduced to calculating the {\em rms} fluctuation.

The primordial gravitational potential is obtained by the
convolution of a real function $\tau({\bf x})$ with a random field
$\varphi({\bf x})$,
$$
\Phi({\bf x}) = \int d^3 y \ \tau({\bf y} - {\bf x}) \varphi({\bf y}).
\eqno(2.4)
$$
The field $\varphi$ is obtained by a non--linear transformation on
a zero--mean Gaussian process $w$, with unit variance and flicker--noise
power--spectrum; the function $\tau$ is fixed by its
Fourier transform,
$$
\tilde{\tau} ({\bf k}) \equiv \int d^3 x e^{-i {\bf k}
\cdot {\bf x}} \tau({\bf x}) = T(k) F(k),
\eqno(2.5)
$$
where $T(k)$ is the CDM transfer function of Eq. (2.2) and $F(k)$ a
positive correction factor which we applied to have the exact CDM initial
power--spectrum of Eq. (2.1) in all our models. The precise form
of the non--linear transformations from $w$ to $\varphi$ are
$ \varphi({\bf x}) \propto e^{w({\bf x})}$
and $ \varphi({\bf x}) \propto w^2({\bf x})$
for $LN$ and $\chi^2$ respectively (Coles \& Barrow 1987; Coles
\& Jones 1991; MMLM).

As MMLM have shown,
both the clustering dynamics and the present large--scale structure
depend strongly upon the sign of the primordial skewness:
positive models rapidly cluster to a lumpy structure with small
coherence length,
while negative models build up a cellular structure
by the slow process of merging of shells around primordial underdense
regions, with larger coherence length. The general conclusion of these
previous studies is that, of the non--Gaussian alternatives considered,
the skew--negative models are the more successful at reproducing the
observed properties of the large--scale structure. Indeed, CMPLMM showed
that very strong constraints on non--Gaussian models of the types
considered here can be placed by the topology test: only Gaussian and
skew--negative models survive the rigours of such an analysis.

\section{The Skewness Test}

CF91 describe the physical motivation behind the use of the skewness of
cell--counts as a diagnostic of large--scale structure, so we just
outline the basics here. Consider the density contrast smoothed on a
certain scale: $\delta_{M}(R)$. In terms of the distribution of
$\delta_{M}(R)$, called $f_{R}(\delta_M)$, we
can define moments as follows:
$$
\langle \left[\delta_{M}(R)\right]^{n} \rangle = \int f_{R}(\delta_M)
\delta_{M}(R)^{n} \, d\delta_{M}(R). \eqno (3.1)
$$
Clearly $\langle \delta_M \rangle =0$;
the quantity $\langle \delta_{M}(R)^{2} \rangle =
\sigma_M^{2}(R)$ is the {\em variance} of
the smoothed mass density fluctuations and $\langle
\delta_{M}(R)^{3} \rangle$ is the
{\em skewness},
denoted $\gamma_{M}(R)$. CF91 found, using a variety of methods in
both the quasi--linear and strong clustering regimes, that for initially
Gaussian density perturbations, $\gamma_M$ grows according to
$$
\gamma_{M}(R) \simeq S\, \sigma_M^{4}(R),\eqno(3.2)
$$
where $S\simeq 3$ is roughly constant, i.e. almost
independent of the scale upon
which $\delta_M$ is smoothed, the background cosmology and the
power--spectrum of the primordial fluctuations.

Since the publication of CF91, many other others have discussed
properties of the skewness and related higher--order moments
(Silk \& Juszkiewicz 1991; Martel \& Freudling 1991; Park 1991;
Lahav {\em et al.} 1992;
Juszkiewicz {\em et al.} 1992; Bouchet {\em et al.} 1992). These other
analyses have confirmed the main conclusions of CF91, but have
demonstrated that $S$ is actually a weak function of $R$ and the
primordial power spectrum. The errors introduced by taking $S$ to be
constant are, however, much smaller than the sampling errors for any
real distribution so we shall ignore these refinements here.
Moreover, all our non--Gaussian models have the same primordial
power--spectrum so we can ignore any dependence on the spectral shape in
this particularly paper.
Bouchet {\em et al.} (1992) have also shown that the effects of
redshift space distortion -- disregarded by CF91 -- should be weak
in the interesting regime. The test proposed by CF91, that one should
plot $\gamma$ against $\sigma^{4}$ for different smoothing scales and
look for departures from linearity, is therefore confirmed as being
a potentially powerful test of non--Gaussian primordial density
fluctuations; CF91 found the QDOT data of Saunders {\em et al.}
(1991) to be consistent with Gaussian initial fluctuations and may be
sensitive enough to rule out viable non--Gaussian scenarios.
Indeed, Silk \& Juszkiewicz (1991) show that
$\gamma \propto \sigma^{3}$ for the cosmic textures model which seems
to be incompatible with the QDOT data. We shall see whether the
non--Gaussian models described in Section 2 are also incompatible with
the data.

There are two main problems when it comes to applying this
test in practice. First, the discrete nature of number--counts of
galaxies itself introduces a skewness term into the cell--count
distribution. Provided that one accepts that the galaxy counts
correspond to a Poisson `shot--noise' effect, then one can easily correct
for the discreteness terms (see below for a discussion). Secondly,
most models of galaxy
formation involve some degree of bias in the ratio of luminous
galaxies to mass. An arbitrary functional bias of the form discussed by
Coles (1992) could seriously interfere with the skewness test. CF91
showed using $N$--body simulations that the standard ``high--peak''
biasing scenario does in fact produce a galaxy distribution with
second-- and third--order moments scaling in the same way as (3.2).
 Nevertheless, different biasing models might produce different
behaviours since an arbitrary bias is in some senses equivalent
to having non--Gaussian fluctuations.

To check the power of the skewness test
in the light of these difficulties, we shall
compare the effects of gravitational evolution on the skewness
of the Gaussian and non--Gaussian models described in Section 2, for
different clustering amplitudes and different levels of bias.

\section{Results}

It is a relatively straightforward matter to extract estimates,
$\hat{\Gamma}(R)$ and $\hat{\Sigma}^{2}(R)$, of
the skewness and variance of cell--counts in cells
of different size $R$ from the simulations described in Section 2.
For large number, $N$, of cells we have
$$
\hat{\Sigma}^{2}  =  \frac{1}{N} \sum_{i=1}^{N} (n_{i}-\bar{n})^{2}
\eqno(4.1a)
$$
and
$$
\hat{\Gamma}  =  \frac{1}{N} \sum_{i=1}^{N} (n_{i}-\bar{n})^{3}
\eqno(4.1b)
$$
where $n_{i}$ is the number of particles in $i$--th cell
and $\bar{n}$ is the mean number per cell, i.e.
$\bar{n}= \sum_{i=1}^{N} n_i/N$. Cell--counts are inevitably skewed
by virtue of their discrete
(integer--valued) nature. To correct for the discreteness terms,
one usually subtracts off `Poisson' terms from the estimates, to
give estimates of $\gamma$ and $\sigma^{2}$ which refer to a
continuously--distributed variable:
$$
\hat{\sigma}^{2}  =  \hat{\Sigma}^{2} - \frac{1}{\bar{n}};\eqno(4.2b)
$$
$$
\hat{\gamma}  =  \hat{\Gamma} - \frac{3\hat{\sigma}^{2}}{\bar{n}}
- \frac{1}{\bar{n}^{2}}\eqno(4.2b)
$$
(Peebles 1980; Saunders {\em et al.} (1991); CF91). Possible reasons why
this might not be an appropriate scheme, particularly if galaxy formation
is in some sense co--operative, are discussed by CF91 (see also Bower
{\em et al.} 1992). To check whether this correction preserves the shape
of the $\gamma$--$\sigma^{2}$ relation (3.2), we look at both
$\hat{\gamma}$--$\hat{\sigma}^{2}$ (i.e. corrected) and
$\hat{\Gamma}$--$\hat{\Sigma}^{2}$  (i.e. uncorrected) relationships.
We need to assign confidence limits to our estimates in order
to assess the significance of departures from the predicted
behaviour. Approximate methods for placing error limits on the
empirically--determined
estimates are discussed by Saunders {\em et al.} (1991), but these
involve a complicated iterative procedure involving high--order
moments. We can make a rough estimate of the error following Kendall
\& Stuart (1977). Suppose each simulation consists of a random
sample of $N$ taken from a Gaussian parent distribution
with variance $\sigma^{2}$. The variance
from sample to sample of estimates of $\gamma$ and $\sigma^{2}$
in such a case are $6\sigma^{6}/N$ and $2\sigma^{4}/N$ respectively.
If we take $\sigma^{2} \simeq \hat{\sigma}^{2}$ then we can
place approximate standard errors, $s$,
on the estimates $\hat{\gamma}$
and $\hat{\sigma}^{2}$ as
$$
s(\hat{\gamma}) \simeq \hat{\sigma}^{2} \sqrt{\frac{6}{N}};\eqno(4.3a)
$$
$$
s(\hat{\sigma}^{2}) \simeq \hat{\sigma}^{2} \sqrt{\frac{2}{N}}.\eqno(4.3b)
$$
Of course, our simulations are not random samples from a Gaussian parent:
the cells are correlated and the distributions are non--Gaussian. We have
also taken the sample variance and the  parent population variance to
be identical. The estimates (4.3) can be expected to give only a rough
order--of--magnitude estimate of the likely confidence limits. We can
also calculate error limits by using the two different
simulations of each model to calculate an estimate of the ensemble
variance analytically. This is still not completely satisfactory --
ideally we would wish for
many more simulations -- but gives results in reasonable accord
with the analytic estimates. We find that the estimated errors
(4.3a,b) exceed the spread of the simulations on large scales by
about 50 per cent, whereas the two estimates agree on intermediate scales.

The simulations also allow us to investigate: (i) the effect of
the normalisation of the model upon the skewness--variance relationship;
(ii) whether the redshift space relationship is significantly
different to that in real space; (iii) whether the relationship
for galaxies identified in the manner described in Section 2 is
different to that of the dark matter particles (iv) whether the
observed QDOT points rule out any of these models.
Our results are displayed in Figures 1--3.

\begin{center}
Figures 1a \& 1b
\end{center}

In Figure 1a we show
the $\hat{\Gamma}$--$\hat{\Sigma}^{2}$ (i.e. uncorrected)
relationship for dark
matter particles (left) and galaxies (right) for all five models
at the {\em present time}. The error bars
are estimated using eq. (4.3a,b). First, note that for
the Gaussian model the form (3.2) is well obeyed for both
the dark matter and the galaxies, as found by CF91. The exception is
at very small scales, where $\sigma^{2}$ is very large. On such small
scales our particle code does not describe the fluid nature of the matter very
well and this results in large discreteness effects (the cell--size here is
comparable to the mesh size). Note that the positively--skewed models
have a systematically higher $\Gamma$ for the same $\Sigma^{2}$;
this is expected because they have higher initial skewness and gravity
acts so as to increase the skewness further. The
trend is somewhat less clear for the negative models. All
such models have positive skewness even to very large scales, which
shows that even weakly non--linear gravitational effects
can wipe out the initial negative skewness. These models, however,
need to be evolved for a comparatively long time in order to
reach the present time. The result seems to be a much smaller systematic
departure from the Gaussian expectation than for the positive--skew
models. Figure 1b
shows the effect of using the corrected values; we show the
$\hat{\gamma}$--$\hat{\sigma}^{2}$ version. The trends are the same
and the quantitative agreement is good, particularly on large scales
where the shot--noise terms are small anyway. Note, however, that on small
scales the shot--noise correction produces a negative skewness (indicated
by the downward arrow plotted at the position of the uncorrected skewness).
This confirms our argument that the failure of the relationship (3.2)
in these simulations is due to a resolution effect.

\begin{center}
Figures 2a \& 2b
\end{center}

To consider the effect of evolution we plot, in Figures 2a and 2b,
corresponding diagrams for models all normalised at $b=1$
(2a) and $b=2.5$ (2b). Only the uncorrected skewness and variance
are shown; discreteness effects act similarly on all our simulations.
Note that the Gaussian model still follows the relationship (3.2)
accurately regardless of the value of $b$; when $b=2.5$ the skewness
on the very largest scale comes out negative but is consistent with
zero within the errors. The positively skew models look closer to the
Gaussian relationship for smaller $b$. The discrepancies for the
negatively--skew models occur in a rather less predictable pattern
and the systematic shift is less apparent.

\begin{center}
Figure 3
\end{center}

To check the effect of redshift distortions we plotted both the real
space and redshift space skewness and variance for all the models
at the present time. The results are plotted in Figure 3. It is clear
that, with the exception of very small scales where resolution effects
dominate, the effect of looking at redshift space rather than real
space is minimal.

We can now look at the question of whether the QDOT points place
any strong constraints on any of these models. These points are
plotted on Figures 1,2 and can be seen to lie on the expected
trajectory (3.2). Although these
points are on large scales from  an observational point of view, they
are on quite small scales compared to these simulations. Looking
only at the corrected results (1b) -- because the QDOT points are
in corrected form --  we see that the two positive--skew models are
clearly excluded at $>2\sigma$; the two negative--skew models
are, however, consistent with the observed points (as is the Gaussian).

\section{Discussion \& Conclusions}

We have investigated the behaviour of the skewness and variance
of the distribution of both dark matter and galaxies in a number
of models involving non--Gaussian initial conditions.

As expected, we find that non--linear gravitational evolution always
acts in such a way that the skewness increases with time. This means
that models with negative initial skewness display a positive skewness
on cosmologically interesting scales even after very weak evolution.
It is difficult therefore to see directly the sign of the initial
skewness. Nevertheless, the fact that models with different initial
skewness evolve in different ways means that their systematic trends
in the behaviour of skewness against variance do remain even into
the fairly strongly non--linear regime. In particular, initially
positive--skew models seem to obey a similar scaling law to the
initially Gaussian models (3.2) but with a higher value of $S$.
For initially negative--skew models, the situation is somewhat
less clear because systematic trends from the Gaussian are smaller.
Part of the reason for this must be that to get models in reasonable
accord with observations on small scales, the negative--skew models
must be highly evolved whereas the positive--skew models are less
strongly evolved (MMLM). This difference in normalisation to the
present epoch tends to suppress differences compared to the Gaussian;
the extra evolution required by the negative--skew models
generates enough skewness to bring them roughly onto the Gaussian
locus in the $\gamma-\sigma^{2}$ plane. Some systematic differences
do remain, especially on large scales, but these are generally
so small as to make discriminating between models difficult.
Moreover, the typical errors for the negative--skew
models are somewhat
larger than those of the positive--skew models. The reason for this
is probably that clustering evolves in the negative--skew models
by forming a quasi--cellular network of bubbles characterised by
a large coherence length. Since there are relatively few of these
structures in the simulations, their presence can produce large
fluctuations from simulation to simulation. Structures in the
positive--skew models generally have a significantly smaller coherence length
and each simulation therefore contains more `typical' structures
than the negative--skew case and fluctuations are correspondingly less.

We have confirmed the conclusion of CF91, that the relationship
(3.2) seems to be reasonably well obeyed for Gaussian models by
both the dark matter and the `galaxy' distributions, at least for
our particular scheme for identifying galaxies. Generally speaking
the behaviour of galaxy and mass fluctuations is similar for all
our models; the most noticeable effect is that in the negative--skew
models, the galaxy skewness lies closer to the Gaussian locus than
the mass fluctuations. The fact remains, however, that such analyses
which rely only on galaxy clustering to test primordial fluctuations
do rely on a particular relationship between galaxies and dark matter.
Complicated non--linear and/or non--local biasing schemes could produce
very different behaviour to that described here (Coles 1992; Bower
{\em et al.} 1992).

By looking at the distribution of matter and galaxies in both
redshift space and real space, we have confirmed that the effect
of peculiar velocity distortions on the relationship (3.2) is small
for all our models. This confirms the argument given by Bouchet
{\em et al.} (1992).

In comparison with the QDOT results for skewness discussed by
Saunders {\em et al.} (1991) we find that initially positive--skew
models fare rather poorly and are excluded by $>2\sigma$. Because
the negative--skew models have rather larger errors associated with
them and the systematic departures from the Gaussian form are
rather smaller than the positive--skew cases, these models cannot be
ruled out by the available skewness measurements. To constrain
these models more strongly, we would need measurements of skewness
with much smaller errors (i.e. from catalogues containing more galaxies)
and preferably out to larger scales.

Of course we must stress that we have considered only a very small subset
of the space of possible non--Gaussian models. All our models have the
CDM power spectrum. Models with more (or less) large (or small) scale
power may well behave differently. We have also chosen models with a
very particular form of statistical distribution, obtained by locally
transforming a Gaussian field.
Galaxy and structure formation
involves a complicated interaction between primordial conditions and
non--linear gravitational evolution and it would be surprising if the
effects of the primordial spectrum and statistics upon the final
density distribution could be separated out completely. We
suspect however -- and other work seems to confirm this idea
(Lahav {\em et al.} 1992; Bouchet {\em et al.} 1992) -- that the
{\em dominant} influence on the skewness--variance relationship
at late times is indeed the primordial skewness of the density
fluctuations. We have not proved this to be true and to do so would
require us to investigate all types of plausible initial power spectra,
background cosmologies and so on. Although our work is thus,
in a sense, limited
it does demonstrate that the skewness of observed galaxy fluctuations is
a potentially powerful probe of the initial distribution of density
fluctuations.

\vspace{1cm}
\noindent{\bf Acknowledgments}

\noindent PC acknowledges the SERC for support under
the QMW rolling theory grant (GR/H09454) and thanks the
Dipartimento di Astronomia at the Universit\`{a} di Padova
for its hospitality during the visit when this work was initiated.
LM acknowledges the Astronomy Centre at the University of Sussex for
the hospitality during the visit when part of this work was done.
FL, SM, AM and LM thank the Ministero Italiano dell'Universit\`{a} e
della Ricerca Scientifica e Tecnologica for financial support.
This work has been partially supported by Consiglio Nazionale delle Ricerche
({\em Progetto Finalizzato: Sistemi Informatici e Calcolo Parallelo}).
The staff and the management of the CINECA Computer Center are warmly
acknowledged for their assistance and for allowing the use of computational
facilities.

\newpage
\large
\begin{center}
\noindent {\bf References}
\end{center}
\normalsize

\begin{trivlist}
\item[] Abbott L.F., Wise M.B., 1984, Nucl. Phys., B244, 541
\item[] Allen T.J., Grinstein B., Wise M.B., 1987, Phys. Lett.,
B197, 66
\item[] Babul A., White S.D.M., 1991, MNRAS, 253, 31P
\item[] Bardeen J.M., Bond J.R., Efstathiou G., 1987, ApJ, 321, 28
\item[] Barrow J.D., Coles P., 1990, MNRAS, 244, 188
\item[] Batuski D.J., Melott A.L., Burns J.O., 1987, ApJ, 322, 48
\item[] Bond J.R., Efstathiou G., Lubin P.M., Meinhold P.R., 1991,
Phys. Rev. Lett., 66, 2179
\item[] Bouchet F.R., Hernquist L., 1992, ApJ, 400, 25
\item[] Bouchet F.R., Juszkiewicz R., Colombi S., Pellat R., 1992,
ApJ, 394, L5
\item[]  Bower R.G., Coles P., Frenk C.S., White S.D.M., 1992, ApJ, 000, 000
\item[] Cen R.Y., Ostriker J.P., Spergel D.N., Turok N., 1991, ApJ, 383, 1
\item[] Coles P., 1992, MNRAS, submitted
\item[] Coles P., Barrow J.D., 1987, MNRAS, 228, 407
\item[] Coles P., Frenk C.S., 1991, MNRAS, 253, 727 (CF91)
\item[] Coles P., Jones B.J.T., 1991, MNRAS, 248, 1
\item[] Coles P., Moscardini L., Plionis M., Lucchin F.,
Matarrese S., Messina A., 1992, MNRAS, 000, 000 (CMPLMM)
\item[] Davis M., Efstathiou G., Frenk C.S., White S.D.M., 1985, ApJ,
292, 371
\item[] Davis R.L., Hodges H.M., Smoot G.F., Steinhardt P.J.,
Turner M.S., 1992, Phys. Rev. Lett., 69, 1856
\item[] Efstathiou G., Kaiser N., Saunders W., Lawrence A.,
Rowan--Robinson M., Ellis R.S., Frenk C.S., 1990, MNRAS, 247, 10P
\item[] Fan Z.H., Bardeen J.M., 1992, preprint
\item[] Ikeuchi S., 1981, PASJ, 33, 211
\item[] Juszkiewicz R., Bouchet F., Colombi S., 1992, ApJ, 000, 000
\item[] Kendall M., Stuart A., 1977,
{\em The Advanced Theory of Statistics},
Volume 1, 4th Edition, Griffin \& Co, London, pp.  257--258
\item[] Kofman L., Blumenthal G., Hodges H., Primack J., 1991,
in Latham D.W. \& da Costa L.N. eds,
{\em Proceedings of the Workshop on Large--Scale Structure and Peculiar
Motions in the Universe}, pp. 339--351, {\em ASP} Conference Series
\item[] La D., Steinhardt P.J., 1989a, Phys. Rev. Lett., 62, 376
\item[] La D., Steinhardt P.J., 1989b, Phys. Lett., B220, 375
\item[] Lahav O., Itoh M., Inagaki S., Suto Y., 1992, ApJ, 000, 000
\item[] Liddle A.R., Lyth D., 1992, Phys. Lett., B291, 391
\item[] Liddle A.R., Wands D., 1991, MNRAS, 253, 637
\item[] Lidsey J.E., Coles P., 1992, MNRAS, 258, 57P
\item[] Lucchin F., Matarrese S., 1985a, Phys. Rev. D, 32, 1316
\item[] Lucchin F., Matarrese S., 1985b, Phys. Lett., B164, 282
\item[] Lucchin F., Matarrese S., Mollerach S., 1992, ApJ Lett., 000, 000
\item[] Maddox S.J., Efstathiou G., Sutherland W.J., Loveday J.,
1990, MNRAS, 242, 43P
\item[] Martel H., Freudling W., 1991, ApJ, 371, 1
\item[] Matarrese S., Lucchin F., Messina A., Moscardini L., 1991,
MNRAS, 252, 35
\item[] Matarrese S., Ortolan A., Lucchin F., 1989, Phys. Rev. D.,
40, 290
\item[] Melott A.L., 1990, Phys. Rep., 193, 1
\item[] Messina A., Lucchin F., Matarrese S., Moscardini L., 1992,
Astroparticle Phys., 000, 000
\item[] Moscardini L., Matarrese S., Lucchin F., Messina A., 1991,
MNRAS, 248, 424 (MMLM)
\item[] Ostriker J.P., Cowie L.L., 1981, ApJ, 243, L127
\item[]  Park C., 1991, ApJ, 382, L59
\item[]  Park C., Spergel D.N., Turok N., 1991, ApJ, 372, L53
\item[] Peacock J.A., 1991, MNRAS, 253, 1P
\item[] Peebles P.J.E., 1980, {\em The Large Scale Structure of the Universe},
Princeton University Press, Princeton
\item[] Salopek D.S., 1992a, Phys. Rev. D., 45, 1139
\item[] Salopek D.S., 1992b, Phys. Rev. Lett., 000, 000
\item[] Salopek D.S., Bond J.R., 1991, Phys. Rev. D., 43, 1005
\item[] Salopek D.S., Bond J.R., Bardeen J.M., 1989,
Phys. Rev. D., 40, 1753
\item[] Saunders W., {\em et al.}, 1991, Nat, 349, 32
\item[] Scherrer R.J., 1992, ApJ, 390, 330
\item[] Scherrer R.J., Bertschinger E., 1991, ApJ, 381, 349
\item[] Silk J., Juszkiewicz R., 1991, Nat, 353, 386
\item[] Smoot G.F., {\em et al.}, 1992, ApJ, 396, L1
\item[] Souradeep T., Sahni V., 1992, MNRAS, 000, 000
\item[] Turok N., 1989, Phys. Rev. Lett., 63, 2625
\item[] Vittorio N., Matarrese S., Lucchin F., 1988, ApJ, 328, 69
\item[] Vittorio N., Meinhold P.R., Muciaccia P.F., Lubin P.M.,
Silk J., 1991, ApJ, 372, L1
\item[] Weinberg D.H., Cole S., 1992, MNRAS, 000, 000
\item[] Weinberg D.H., Ostriker J.P., Dekel A., 1989,
ApJ, 336, 9
\end{trivlist}
\newpage
\begin{center}
\noindent {\bf Figure Captions}
\end{center}

\vspace{1cm}
\noindent 1. The relationship between skewness, $\gamma$, and variance,
$\sigma^{2}$, for our five models. Figure 1a shows the results without
shot--noise correction; Figure 1b has shot--noise corrections.
The solid line is the theoretical relationship (3.2). The
downward--pointing arrows in 1b indicate that the corrected skewness
is negative; the arrows are plotted at the corrected variance
value and originate at the uncorrected skewness value. All models
are normalised to the present time (see Section 2). The three crosses
are the QDOT measurements with error bars. Error bars  on the simulated
results are estimated using eq. (4.3a,b).

\vspace{1cm}
\noindent 2. The effect of normalisation upon the skewness--variance
relationship. Figure 2a shows all models normalised to $b=1$ and
Figure 2b shows the normalisation $b=2.5$.
The solid line is the theoretical relationship (3.2).
All points are uncorrected for shot--noise. The small downward
arrows on the theoretical line indicate that the uncorrected
skewness is negative at that point. The three crosses are the QDOT
measurements with error bars. Error bars on the simulated results
are estimated using eq. (4.3a,b).

\vspace{1cm}
\noindent 3. Skewness--variance relationships in real space and velocity
(redshift) space. Filled circles show $\gamma$ and $\sigma^{2}$ in real
space, open circles show velocity space. The two circles generally coincide
when $\sigma^{2}$ is small. All models are normalised to the present time
and we plot the shot--noise corrected results. Error bars
are estimated using eq. (4.3a,b).

\end{document}